\documentclass[sigconf,10pt]{acmart}
\usepackage{soul}
\usepackage{subfigure}
\usepackage{microtype}

\AtBeginDocument{%
  \providecommand\BibTeX{{%
    \normalfont B\kern-0.5em{\scshape i\kern-0.25em b}\kern-0.8em\TeX}}}

\setcopyright{none}

\renewcommand\footnotetextcopyrightpermission[1]{} 

\acmConference[C4ML '20]{C4ML '20: Compilers for Machine Learning Workshop at CGO 2020}{February 23, 2020}{San Diego, CA}
\acmBooktitle{C4ML '20: Compilers for Machine Learning Workshop at CGO 2020, February 23, 2020, San Diego, CA}

\begin{document}

\title{Optimizing Memory-Access Patterns for Deep Learning Accelerators}

\author{Hongbin Zheng, Sejong Oh, Huiqing Wang, Preston Briggs, Jiading Gai,}
\author{Animesh Jain, Yizhi Liu, Rich Heaton, Randy Huang, Yida Wang}
\affiliation{
  \institution{Amazon Web Services}            
}

\renewcommand{\shortauthors}{Zheng, et al.}

\begin{abstract}
Deep learning (DL) workloads are moving towards \emph{accelerators\/} for faster processing and lower cost. Modern DL accelerators are good at handling the large-scale multiply-accumulate operations that dominate DL workloads; however, it is challenging to make full use of the compute power of an accelerator since the data must be properly staged in a software-managed scratchpad memory. Failing to do so can result in significant performance loss. This paper proposes a systematic approach which leverages the polyhedral model to analyze all operators of a DL model together to minimize the number of memory accesses. Experiments show that our approach can substantially reduce the impact of memory accesses required by common neural-network models on a homegrown AWS machine-learning inference chip named \emph{Inferentia}, which is available through Amazon EC2 Inf1 instances.
\end{abstract}



\maketitle

\section{Introduction}
As deep learning (DL) models grow in sophistication and computational load, the traditional approach of executing DL workloads, \emph{i.e.,} neural networks, on CPUs and GPUs is becoming more time consuming and expensive. There is a trend to move DL workloads to custom accelerators~\cite{jouppi2018domain, chen2016diannao}. By designing domain-specific architectures, these processors are able to accelerate DL workloads and reduce energy requirements by orders of magnitude.

A typical DL model can be represented as a 
graph, where nodes are operators and directed edges denote the dependences between nodes. Modern accelerators mostly focus on compute-bound operators such as \emph{convolution (CONV)\/} and \emph{general matrix multiplication (GEMM)\/} via specially designed compute units like systolic arrays. These units are able to process multiply-accumulate operations in a highly efficient manner. On the other hand, the accelerators depend on complex software-managed scratchpads. End-to-end performance will be limited if memory references of a neural network are not well organized. Current solutions, \emph{e.g.,} the XLA compiler for Google's TPU~\cite{xla}, handle memory-access optimization within an operator, but ignore opportunities to reduce the number of memory accesses across multiple operators. There is some global optimization work for DL models~\cite{jia2019optimizing, liu2019optimizing}, but no one seems to have attacked global optimization of memory-access patterns for DL accelerators.


We propose a systematic way to optimize the memory-access patterns of DL models for efficient execution on DL accelerators. Specifically, our approach takes a DL model as input, does a number of global optimizations to remove unnecessary memory copies and intelligently schedule necessary memory accesses on the accelerators to maximize the memory-bandwidth usage. Experiments show that we are able to significantly reduce the impact of memory references running on a homegrown AWS machine-learning inference chip named \emph{Inferentia}. The chip is available to public through Amazon EC2 Inf1 instances~\footnote{\url{https://aws.amazon.com/ec2/instance-types/inf1/}}.

\section{Optimization Method}

Our work is part of the compiler toolchain for \emph{Inferentia}. The toolchain reads in the computation graph of a DL model, defines the operators via TVM~\cite{chen2018tvm} to build an intermediate representation (IR) that represents the whole neural network, applies analyses and optimizations to the IR, and eventually produces a low-level IR for machine-code generation.

This paper focuses on a small portion of the compiler: optimizing the memory-access patterns. A DL workload manipulates high dimensional tensors with \emph{loop nests}. Without loss of generality, we define the tensor accesses with element-wise load and store instructions inside a loop nest based on the polyhedral model~\cite{verdoolaege2016presburger}:
\begin{align} \label{eqn:load-inst}
\tag{load}
v_l = t_m[\vec{f}(\vec{i})]
\end{align}
\vspace{-2em}
\begin{align} \label{eqn:store-inst}
\tag{store}
t_m[\vec{f}(\vec{i}))] = v_s
\end{align}
In these definitions, $\vec{i}=i_0,i_1,...,i_{n-1}$ represents a loop nest with $n\/$ loops, where $i_j$ is the loop at level $j$,
$t_m$ represents the $m\/$-dimensional tensor which is being read/written by the load/store instructions, and $\vec{f}(\vec{i}) =
C\vec{i} + \vec{b}$.
Since the matrix $C\/$ and the vector $b\/$ are compile-time constants, $C\vec{i}+\vec{b}$
is an \emph{affine\/} expression.
Finally, $v_l$ in (\ref{eqn:load-inst}) represents the result of the load instruction and $v_s$ in (\ref{eqn:store-inst}) represents the data being written to $t_m[\vec{f}(\vec{i}))]$ in the store instruction.

Our approach tries to eliminate unnecessary data movements in the workload (Section~\ref{sec:dme}), and for the remainder, maximizing the utilization of the on-chip memory by maintaining data locality in the scratchpad (Section~\ref{sec:bank-mapping}). Our approach was designed for DL accelerators equipped with powerful compute units and limited on-chip memory.

\subsection{Data-Movement Elimination} 
\label{sec:dme}
Data-movement elimination tries to eliminate the pair of instructions $(v = t_l[\vec{f_l}(\vec{i})], t_s[\vec{f_s}(\vec{i}))] = v)$, where the result of the load instruction, $v$, directly feeds the input of the store instruction. Such patterns are found in DL workloads by analyzing the loop nests of pairs of memory-bound operators like \emph{repeat, tile, split, transpose, strided\_slice, etc.} Existing solutions cannot thoroughly eliminate them without optimizing globally.

To eliminate such pairs, we first generate the reverse of $\vec{f_s}$ as $\vec{f_s'} : \vec{idx}_{t_s} \mapsto \vec{i} $. Using $\vec{f_s'}$, we build a function:
\begin{align} \label{eqn:store-to-load}
\vec{g_{ls}} = \vec{f_l} \circ \vec{f_s'} = \vec{f_l}(\vec{f_s'}(\vec{idx}_{t_s})) : \vec{idx}_{t_s} \mapsto \vec{idx}_{t_l}
\end{align}
to map the index space of tensor $t_s$ to the index space of tensor $t_l$. Using $\vec{g_{ls}}$, we rewrite the accesses that read $t_s$ so they directly read $t_l$ which in turn allows us to eliminate the stores that defined $t_s$. Specifically, for each load instruction that reads $t_s$, $v' = t_s[\vec{f_l'}(\vec{i'})]$, we build a function:
\begin{align} \label{eqn:load-store-to-load}
\vec{g'} = \vec{g_{ls}} \circ \vec{f_l'} = \vec{g_{ls}}(\vec{f_l'}(\vec{i'})) =  \vec{f_l}(\vec{f_s'}(\vec{f_l'}(\vec{i'}))) : \vec{i'} \mapsto \vec{idx}_{t_l}
\end{align}
to map the loop indices $\vec{i'}$ to the index space of $t_l$ and rewrite the load instruction $v' = t_l[\vec{g'}(\vec{i}')]$. Once we apply such transformations to all load instructions that read tensor $t_s$, $t_s$ can be eliminated along with all instructions defining it.

We repeat this process until we cannot eliminate any more load/store pairs. The affine function \emph{reverse\/} and \emph{composition\/} are implemented using the Integer Set Library~\cite{verdoolaege2010isl}.


\subsection{Global Memory-Bank Mapping}
\label{sec:bank-mapping}
Not all data movement in a DL workload can be removed. For compulsory references, we try to fully exploit the available memory bandwidth. In order to maximize the internal memory bandwidth, accelerators typically organize on-chip memories into multiple banks with disjoint address spaces, each of which connects to one portion of the compute units (\emph{e.g.,} a specific row of the systolic array). Data movement between different banks is very slow through the main memory; therefore, tensor data needs to be carefully spread across the banks for computation. For example, in a \emph{Conv2D\/} operator, data from different channels of the feature map and weights must be mapped to different memory banks so that the internal compute units can read and process the data in parallel. At the same time, the result of the \emph{Conv2D\/} needs to be spread across several banks, guided by the different output channels. 

In prior work~\cite{ding2014blp}, bank mapping focused on a single loop nest with a goal of maximizing the memory-access parallelism for that nest. We call this \emph{local bank mapping.}

Our goal is to minimize inter-bank data movement between multiple operators (represented by multiple loop nests in our compiler).
To achieve this goal, we first derive bank mappings for the operators with bank-mapping restrictions, \emph{e.g., conv2D, matmul, pooling, etc.,} then propagate these mappings across the network based on the data dependencies between operators.
We perform a fixed-point iteration to propagate the mappings to cover all operators in the neural network and make sure that the output of an operator maps to the memory banks required by the next operator. 
If a tensor $t\/$ has conflicting mapping requirements during the propagation, \emph{i.e.,} the data layout changes between consecutive operators in the network, we will introduce a tensor $t'$ and a \emph{memcopy\/} between $t\/$ and $t'$ to represent data movement between memory banks.
Typically, for a high-dimensional tensor, we map its outer dimensions to different banks and use its inner dimensions to address different elements in the same bank to support sequential data access.





\section{Evaluation}

We conducted our experiments on a homegrown AWS chip called \emph{Inferentia}, specifically, Amazon EC2 Inf1.xlarge instance. For the sake of space, we present results of a single model for each algorithm.

We tested the effectiveness of data-movement elimination on Parallel WaveNet~\cite{oord2017parallel}. Our optimization was able to eliminate 123 out of 124 load-store pairs. As a result, we eliminated 145~MB (out of 146~MB) 
of tensors that were used for intermediate storage. We saved 10\% of the on-chip memory copies and 11\% of the off-chip memory copies (measured in bytes).

We tested the effectiveness of global memory-bank mapping by running our compiler on ResNet-50~\cite{he2016deep}, comparing two different mapping algorithms:
\begin{description}
\item[Local mapping] which generates mappings within each operator, without propagation, but keeps the output of an operator in on-chip memory if it will be directly used as the input of the next operator.
\item[Global mapping] as described in Section~\ref{sec:bank-mapping}.
\end{description}
Taking results from local mapping as a baseline, we saw global mapping eliminate 76\% of the on-chip data copies and 37\% of the copies off chip (measured in bytes).

\section{Conclusion}
To conclude, this paper proposes a systematic approach to globally optimize the memory-access patterns of DL workloads on accelerators. Experimental results show that we are able to
significantly reduce memory references for state-of-the-art networks on \emph{Inferentia,} a homegrown AWS machine-learning inference chip.

\bibliographystyle{ACM-Reference-Format}
\bibliography{ref}

\end{document}